\documentclass[10pt]{amsart}
\usepackage{amsmath}
\usepackage{amssymb}
\usepackage{amsthm}

\newcommand{\RR}{\mathbb{R}}

\newcommand{\NN}{\mathbb{N}}
\newcommand{\ZZ}{\mathbb{Z}}
\DeclareMathOperator{\EE}{\mathbb{E}}

\newcommand{\cB}{\mathcal{B}}
\newcommand{\cE}{\mathcal{E}}
\newcommand{\cJ}{\mathcal{J}}
\newcommand{\cN}{\mathcal{N}}

\renewcommand{\l}{\left}

\DeclareMathOperator{\sign}{\mathrm{sign}}
\newcommand{\eff}{\mathrm{eff}}
\newcommand{\Erw}{ \mathbf{E}_{x} }
\newcommand{\Prb}{ \mathbf{P}_{x} }
\newcommand{\Tr}{{\mathop{\mathrm{Tr}}}}

\newcommand{\loc}{{\mathop{\mathrm{loc}}}}
\newcommand{\per}{{\mathop{\mathrm{per}}}}
\DeclareMathOperator{\supp}{\mathrm{supp}}
\DeclareMathOperator{\diam}{\mathrm{diam}}
\newtheorem{thm}{Theorem}%[section]
\newtheorem{lem}[thm]{Lemma}

\theoremstyle{definition}

\newcommand{\Hmm}[1]{\leavevmode{\marginpar{\tiny%
$\hbox to 0mm{\hspace*{-0.5mm}$\leftarrow$\hss}%
\vcenter{\vrule depth 0.1mm height 0.1mm width \the\marginparwidth}%
\hbox to 0mm{\hss$\rightarrow$\hspace*{-0.5mm}}$\\\relax\raggedright #1}}}
\begin{document}
\title[Spectral shift and Wegner estimates]
{Bounds on the spectral shift function and the density of states}
\author[D.~Hundertmark]{Dirk Hundertmark}
\author[R.~Killip]{Rowan Killip}
\author[S.~Nakamura]{Shu Nakamura}
\author[P.~Stollmann]{Peter Stollmann}
\author[I.~Veseli\'c]{Ivan Veseli\'c}
\address[D.~Hundertmark]{Department of Mathematics, University of Illinois at Urbana-\-Champaign,
 Illinois 61801, USA} \email{dirk@math.uiuc.edu} \urladdr{www.math.uiuc.edu/\protect{\symbol{20}}dirk/}
\address[R.~Killip]{Mathematics Department, UCLA, Los Angeles, CA 90095-1555, USA}
\email{killip@math.ucla.edu} \urladdr{www.math.ucla.edu/\protect{\symbol{20}}killip/}
\address[S.~Nakamura]{Graduate School of Mathematical Science, University of Tokyo, 3-8-1 Komaba,
 Meguro Tokyo, 153-8914 Japan} \email{shu@ms.u-tokyo.ac.jp} \urladdr{www.ms.u-tokyo.ac.jp/\protect{\symbol{20}}shu/}
\address[P.~Stollmann]{Technische Universit\"at Chemnitz, Fakult\"at f\"ur Mathematik, D-09107 Chemnitz, Germany}
\curraddr{} \email{P.Stollmann@mathematik.tu-chemnitz.de} \urladdr{www.tu-chemnitz.de/\protect{\symbol{20}}stollman/}
\address[I.~Veseli\'c]{ Department of Mathematics, California Institute of Technology, USA }
\curraddr{Technische Universit\"at Chemnitz, Fakult\"at f\"ur Mathematik,
            D-09107 \mbox{Chemnitz}, Germany}
\email{ivan.veselic@mathematik.tu-chemnitz.de}
\urladdr{www.tu-chemnitz.de/mathematik/schroedinger/index.php}
\date{\today, \jobname.tex}

\keywords{integrated density of states, random Schr\"odinger
operators, alloy type model} \subjclass[2000]{35J10,81Q10}

\begin{abstract}
We study spectra of Schr\"odinger operators on $\RR^d$. First we consider a pair of operators which differ
by a compactly supported potential, as well as the corresponding semigroups. We prove almost exponential decay
of the singular values $\mu_n$ of the difference of the semigroups as $n\to \infty$ and
deduce bounds on the spectral shift function of the pair of operators.

Thereafter we consider alloy type random Schr\"odinger operators. The single site potential $u$ is assumed to
be non-negative and of compact support. The distributions of the random coupling constants are assumed to be
H\"older continuous. Based on the estimates for the spectral shift function, we prove a Wegner estimate which
implies H\"older continuity of the integrated density of states.
\end{abstract}

\maketitle

\section{Introduction and Results}\label{s-results}

In this paper we analyze the spectral properties of multi dimensional Schr\"odinger operators.
First, we consider a pair of operators $H_1,H_2$ which differ by a compactly supported potential $u$.
The singular values $\mu_n$ of the difference of the corresponding exponentials $V_{\eff}:=e^{-H_1}- e^{-H_2}$
are shown to decay almost exponentially in $n$.

This result allows us to deduce a bound on the Lifshitz-Krein   spectral shift function (SSF) of the operator
pair $H_1,H_2$. We give a bound on the SSF when integrated over the energy axis against a bounded, compactly
supported function. In turn,   the bound on the SSF is used to prove a Wegner estimate for random
Schr\"odinger operators of alloy type and H\"older continuity of the integrated density of states (IDS). Our
estimates have a better continuity in the energy parameter than previously known   bounds. Moreover, we are
able to treat random coupling constants, whose distribution does not have a density. In particular, for
H\"older continuous distributions we prove that the IDS is H\"older continuous, too.

\medskip

We will treat magnetic Schr\"odinger operators
\begin{equation}\label{E:Hdefn}
  H=H_A+V=(-i\nabla-A)^2+V
\end{equation}
acting on $\RR^d$ whose potentials (magnetic and electric) obey the following hypotheses:
Each component of $A$ is $L_{\loc}^2$. The positive part of the electric potential, $V_+ := \max(0,V)$,
belongs to $L_{\loc}^1$ and the negative part, $V_-:= \max(0,-V)$, is in the Kato class.
Notice that under our convention, $V=V_+ - V_-$.

For a general discussion of the Kato-class, see \cite{CyconFKS}; for its relevance to the Feynman-Kac formula
see, e.g., \cite{AizenmanS,BroderixHL,Simon-82}. In particular, $V_-$ is in the Kato-class, if
\begin{displaymath}
\Vert V_- \Vert_{L^p_{\text{loc},\text{unif}}(\RR^d)}
= \sup_{x\in\RR^d} \Big( \int_{|x-y|\le 1} \vert V_-(y)\, dy\vert \Big)^{1/p} <\infty
\end{displaymath}
where $p=1$ if $d=1$ and $p>d/2$ if $d\ge 2$. Thus the allowed potentials 
cover all
physically relevant cases.

Under these hypotheses, one may define $H$ via the corresponding quadratic form
(with core $C^\infty_c$).  By the same method, one can define the Dirichlet restriction of $H$
to the cube $\Lambda_l=[-l/2,l/2]^d,l \ge 1$.  This will be denoted $H^l$.
\smallskip

Let $H_1$ be a Schr\"odinger operator of the form just described and let $H_2=H_1+u$
where $u=u_+ - u_-$ obeys the hypotheses for electric potentials just described.

The starting point for our analysis is an estimate on the singular values of $V_{\eff}:=e^{-H_1}- e^{-H_2}$
and on the corresponding object in the finite volume case, namely $V_{\eff}^l:=e^{-H_1^l}- e^{-H_2^l}$.
Recall that the singular
values of a compact operator $A$ are the square-roots of the eigenvalues of $A^*A$.  We will enumerate them
as $\mu_1(A)\geq \mu_2(A)\geq \cdots\geq 0$ according to multiplicity.

\begin{thm}
\label{t-singval}
There are finite positive constants $c$ and $C$ such that the singular values of the operator $V_{\eff}^l$ obey
\begin{equation}\label{e-singval}
  \mu_n \le C \, e^{-c  n^{1/d}}.
\end{equation}
In fact, $c$ may be chosen depending only on the dimension, while $C$ depends on the Kato-class
norms of $u_-,V_-$ and on the diameter of the support of $u_+$.

The same estimate holds for the singular values of $V_{\eff}$.
\end{thm}

\noindent
\textbf{Remarks:} i)
In particular $\|V_{\eff}^l\|_{\cJ_p}:=\sum_n \mu_n^p $ is finite
and thus $V_{\eff}^l$ is an element of the operator ideal
$\cJ_p:=\{A \text{ compact }| \, \|A\|_{\cJ_p}<\infty\}$ for any
$p>0$. Thus our result can be understood as a sharpening and generalization
of norm, Hilbert Schmidt, and trace bounds on the difference of semigroups, derived
e.g.~in \cite{vanCasterenD-89a,vanCasterenD-94,vanCasterenDSS-95,DeiftS-76,Demuth-80,%
Demuth-99, DemuthKM-93,Simon-79c, Stollmann-94b,Stollmann-94a}.

\smallskip

\noindent
ii) Note that the estimate \eqref{e-singval} depends on the
positive part of $u$ only through $\supp u$. Thus, for $u=\lambda
\tilde u$ where $\tilde u \ge 0$ and $\lambda$ is a non-negative
coupling constant, the estimate is independent of the choice of
$\lambda$. Moreover

(1) $u$ may be taken to $+\infty$ on its support. In this case $H_2$ equals the restriction of $H_1$ to
$\RR^d\setminus \supp \, u$ with Dirichlet boundary conditions, provided the boundary of $\supp \, u$ obeys some mild regularity conditions;
see, e.g., \cite{Stollmann-95}.

(2) Similarly, $H_1$ may be defined on a set strictly smaller than
$\RR^d$: Let $D\subset$ be open, $H_A^D$ the Dirichlet
restriction of $H_A$ on $D$, and $H_1=H_A^D + V$ where $V$
satisfies the same conditions  as before. In this case
$H_j^l$ is the Dirichlet restriction of $H_j, j=1,2$ to the set
$\Lambda_l \cap D$.

\smallskip

\noindent
iii) The proof of Theorem \ref{t-singval} is surprisingly simple. Morally, the result is an
immediate consequence of Weyl's law for the eigenvalue asymptotic of Dirichlet
Laplacians on compact domains. This suggest that the decay rate of the singular values of
$e^{-H_1}- e^{-H_2}$ is, in fact, given by $\exp(-cn^{2/d})$.
One might ask, however, whether the singular values could not typically decay at a much
faster rate. It turns out that a decay rate like $\exp(-c n^\alpha)$ for $\alpha > 2/d$
is impossible, see Remark ii) after Theorem \ref{t-SSF} below. This leaves open the
cases $\alpha \in (1/d, 2/d]$. We conjecture that, in fact, the true bound is of
the form
$
\mu_n\le C \exp(-c n^{2/d}) .
$

\smallskip

Our interest in Theorem \ref{t-singval} comes from the fact that it allows us to derive an integral
bound on the the SSF $\xi(\lambda, H_2, H_1)$ of the pair of operators $H_1,H_2$, which shows that
the SSF can have only very mild local singularities. (See Section \ref{s-DefSSF} for a precise
definition of the SSF.) The SSF plays a role in different areas of mathematical physics, for
instance in scattering theory, cf.~e.g.~\cite{Yafaev-92}, and the study of surface potentials, see
\cite{Chahrour-00,KostrykinS-01b}. Various of its properties are discussed in the literature:
monotonicity and concavity in \cite{GeislerKS-95,GesztesyMM-99,Kostrykin-00}, the asymptotic
behaviour in the large coupling constant \cite{Pushnitski-00,Safronov-01,PushnitskiR-02} and
semiclassical limit \cite{Pushnitski-99,Nakamura-99}. See \cite{BirmanY-93,Kostrykin-99} for  surveys.

For $t>0$ let $F_t:[0,\infty)\to [0,\infty)$ be defined by
\begin{equation}\label{e-definitionF}
F_t(x) = \int_0^x (\exp(t y^{1/d})-1)\, dy.
\end{equation}
As the integrand is increasing, $F_t$ is a convex function.

\smallskip

\begin{thm} \label{t-SSF}
Let $\xi$ be the spectral shift function for the pair $H_1, H_2$ or
$H_1^l, H_2^l$. \\[0.1em]
i{\rm)} Let $F_t$ be defined as above. There exisits a constant $K_1$,
depending on $t$, such that for small enough $t>0$,
\begin{equation}\label{e-integral}
\int_{-\infty}^T F_t(|\xi(\lambda)|)\, d\lambda \le K_1 e^T <\infty
\end{equation}
for all $T<\infty$. \\[0.1em]
ii{\rm)} There exists constants $K_1, K_2$ depending only on $d$, $\diam \supp \, u_+$ and the Kato class norms of $V_-, u_-$,
 such that for any bounded compactly supported function $f$,
\begin{eqnarray}\label{e-dual}
  \int f(\lambda)  \, \xi(\lambda) \, d\lambda
  \le K_1 e^b + K_2 \, \{\log(1+ \|f\|_\infty)\}^d \|f\|_1
\end{eqnarray}
with $b=\sup \supp(f)$.
\end{thm}

\noindent
\textbf{Remarks:} i) Note that the function $F_t$ defined in \eqref{e-definitionF}
has the asymptotic behavior
\begin{displaymath}
F_t(x)\sim d\, x^{(d-1)/d}\exp(tx^{1/d}) \qquad\text{ for large } x.
\end{displaymath}
Thus, by part i) of Theorem \ref{t-SSF}, the spectral shift function can have
at most mild logarithmic local singularities. It is tempting to think that,
at least for non-negative compactly supported perturbations, the spectral shift function
should always be bounded. However, this is not the case. For a perturbation of
the free Schr\"odinger operator with a constant magnetic field by a compactly
supported potential, Raikov and Warzel showed that spectral shift function
diverges at each Landau level $E_q$ like
\begin{equation}\label{e-RWexample}
\vert \xi(E_q+\lambda) \vert
\sim \Big( \frac{|\ln(\lambda)|}{\ln \vert\ln\lambda\vert } \Big)^{d/2}
\qquad \text{ as }\lambda\downarrow 0 ,
\end{equation}
see \cite{RaikovW-02b} for the case $d=2$ and \cite{RozenblumM} for the
generalization to even dimensions.
Thus, setting $F_{t,\alpha}(x):= \int_0^x(\exp(t y^\alpha)-1)\, dy$,
the asymptotic \eqref{e-RWexample} implies that $F_{t,\alpha}(\vert\xi\vert)$ is
locally integrable if and only if $0\le \alpha \le 2/d$, whereas
Theorem~\ref{t-SSF} guarantees it only for $\alpha\le 1/d$ (and $t$ small enough,
if $\alpha = 1/d$).

\smallskip

\noindent ii) The proof of Theorem \ref{t-SSF} shows that if Theorem \ref{t-singval} holds in the
form $\mu_n\le c_1\exp(-c_2 n^\alpha)$, then $F_{t,\alpha}(\vert\xi\vert)\in L^1(-\infty,T)$ for
small enough $t$ (and all finite $T$). Thus the Raikov-Warzel result shows that the estimate in
Theorem~\ref{t-singval} \emph{cannot} be improved beyond $C \exp(-c n^{2/d})$.

\smallskip

\noindent
iii) An example without magnetic fields, where the SSF shows unexpected divergencies, was
studied by Kirsch in \cite{Kirsch-87,Kirsch-89b}. It is related to the one with a
constant magnetic field in that the high degeneracy of eigenvalues plays a crucial role.
Let $E>0$, $u $ non-negative, bounded with compact support, and not identically equal to zero,
$a\colon [0,\infty) \to (0,\infty)$,
and $\xi_l(\cdot):= \xi(\cdot, -\Delta^l, (-\Delta+a(l) u)^l)$.
Then $\limsup_{l\to \infty} \xi_l(E)= \infty$, for any $E, a$ and $u$ as above.
This result relies on the degeneracy of eigenvalues of the pure
Dirichlet Laplacian on a cube. There is, however, a set of full measure
$\cE\subset \RR$ with dense complement such that
$\lim_{\NN \ni l\to \infty} \xi_l(E) = 0$, for all $E \in \cE$, if $a(l) \le l^{-k}, k >3$.

\smallskip

\noindent
iv) In contrast to the above unboundedness results, Sobolev, \cite{Sobolev-93} showed
that for the pair $H_1=-\Delta$ and $H_2= -\Delta +u$ with
$|u(x)|\le const. \, (1+|x|)^{-\alpha}$ and $\alpha >d$, the spectral shift function
$\xi$ is, indeed, locally bounded. However, this type of result seems to require
very strong hypotheses on $H_1$, for example, a limiting absorption principle and in particular, that $H_1$
has absolutely continuous spectrum on the positive real axis.

\smallskip

Theorem \ref{t-SSF}.ii) has a nice consequence in the theory of random Schr\"odinger operators. In
this case, we take $f$ to be the the derivative of a smooth, monotone switch function
$\rho:=\rho_{E,\varepsilon}\colon \RR \to [-1,0]$. \label{p-rho} By a switch function we mean that
for a positive $\varepsilon \le 1/2$ it has the following properties: $ \rho\equiv -1$ on
$(-\infty,E-\varepsilon]$, $\rho\equiv 0$ on $[E+\varepsilon,\infty)$ and $\|\rho'\|_\infty \le
1/\varepsilon$. Theorem \ref{t-SSF}.ii) and the Krein trace identity, see \S\ref{s-DefSSF}, then
imply that there is a constant $C_E$ such that
\begin{equation} \label{e-SSF}
\Tr \left [ \rho (H_2)-\rho (H_1)  \right  ] \le C_E \, |\log( \varepsilon)|^d.
\end{equation}

The estimate \eqref{e-SSF} improves upon the bound derived by Combes, Hislop and Nakamura in \cite{CombesHN-2001}.
They prove that for any exponent $\alpha <1$, there is a constant $\tilde C_E(\alpha)$
depending only on $d$, $C_0$, $\diam \supp u$, $E+\varepsilon$ and $\alpha$ such that
\begin{equation} \label{e-CHN-SSF}
\Tr \left [ \rho (H_2)-\rho (H_1)  \right  ] \le \tilde C_E(\alpha) \ \varepsilon^{-\alpha} .
\end{equation}

An \emph{alloy type model} is a random Schr\"odinger operator
$H_\omega =H_0 +V_\omega$, where $H_0=H_A+V_{\per}$ with a periodic potential $V_\per$.
The random part of the potential has the form
$V_\omega(x)=\sum_{k\in\ZZ^d} \omega_k \, u(x-k)$. The
\emph{coupling constants} $\omega_k, k \in \ZZ^d$, are a sequence
of bounded random variables, which are independent and identically
distributed with distribution $\mu$. The expectation of the product measure
$\bigotimes_{k\in \ZZ^d} \mu$
is denoted by $\EE$. The single \emph{site potential} $u\not\equiv 0$ is
of compact support. Denote for $\varepsilon>0$
\begin{equation}
\label{definition-s-mu-epsilon}
s(\mu,\varepsilon)=\sup\{\mu([E-\varepsilon,E+\varepsilon]) \mid E \in \RR\} .
\end{equation}
With this definition, we have

\begin{thm}
\label{t-WE} Let $H_\omega$ be an alloy type model and $u\ge \kappa \chi_{[-1/2,1/2]^d}$ for some positive
$\kappa$. Then for each $E_0\in \RR$ there exists a constant $C_W$ such that, for all $E\le E_0$ and
$\varepsilon \le 1/2$
\begin{equation}
\label{e-WE}
\EE\{\Tr [ \chi_{[E-\varepsilon,E+\varepsilon]}(H_\omega^l) ]\}
\le C_W \ s(\mu,\varepsilon) \, (\log \textstyle \frac{1}{\varepsilon})^d \ |\Lambda_l|
\end{equation}
\end{thm}

In particular, if $\mu$ is  H\"older continuous with exponent $\alpha$, then the $\varepsilon$-dependence of
the RHS~of \eqref{e-WE}  is $\varepsilon^\alpha  (\log \textstyle \frac{1}{\varepsilon})^d $. In
\cite{Stollmann-00b}, Stollmann proved  a weaker version of \eqref{e-WE} with RHS~equal to $C_W \
s(\mu,\varepsilon) \, |\Lambda_l|^2 $.

Bounds like \eqref{e-WE} are called \emph{Wegner estimates}. They were first deduced by physical reasoning by Wegner
in \cite{Wegner-81} for the \emph{Anderson model}, the finite difference analogue of the alloy type model.
Wegner estimates are important \emph{a priori} estimates, used to derive regularity properties of the integrated
density of states (IDS) and to prove localization for random Schr\"odinger operators. In this context,
localization means the existence of an energy region where the random operator has almost surely dense pure
point spectrum with exponentially decaying eigenfunctions. See, e.g., ~\cite{Stollmann-01} for a monograph
exposition and, e.g.,  \cite{GerminetK-2001b} (and the references therein) for more recent developments.

The proof of Theorem \ref{t-WE} can be directly applied to the Anderson model.
In this case, one obtains a bound like \eqref{e-WE} with the RHS side equal to
$ C_W \ s(\mu,\varepsilon) \, |\Lambda| $ and the constant $C_W$ independent
of the energy $E$. This gives an easy proof of a Wegner estimate for H\"older
continuous single site distributions $\mu$.

The IDS $N(E)$ is defined as the limit of the distribution
functions,
\[
N_\omega^l(E):=|\Lambda_l|^{-1} \ \# \{ \text{ eigenvalues
of }  H_\omega^l \text{  not greater than } E \} ,
\]
as $l$ tends to infinity. For almost all $\omega \in \Omega$ the
limit exists and is independent of $\omega$.
As a consequence of Theorem \ref{t-WE}, the IDS of the above alloy-type
model satisfies
\[
|N(E_1)-N(E_2)| \le C_I \ s(\mu,|E_1-E_2|) \,  \l(\log\frac{1}{|E_1-E_2|}\right)^d,
\quad |E_1-E_2| \le 1/2 ,
\]
where the constant $C_I$ may be chosen uniformly if $E_1$ and
$E_2$ vary in a compact interval $I$. Note that this continuity result cannot
be obtained from a Wegner estimate with quadratic dependence on the volume of
the box $\Lambda_l$. It also shows that, up to a logarithmic correction,
the IDS enjoys the same regularity properties as the distribution of the
random potential.
\smallskip

Let us return to the discussion of the regularity of the SSF. Denote by $H_1=H_\omega - \omega_0 u$
and $H_2=H_1+u$ the alloy type operators where the value of the coupling constant at $k=0$ is frozen
and equal to $0$ and $1$ respectively. The other coupling constants are still random.  Despite the
examples given above it is still possible that the \textit{average} of the SSF
$\bar\xi(\lambda):= \EE\{\xi (\lambda,H_1, H_2 )\}$ over the \emph{random background} environment
is locally bounded. This then implies no logarithmic loss in the Wegner estimate, and thus
the Lipschitz continuity of the integrated density of states.

Indeed, such a bound on $\bar\xi$ for $d \le 3$ has recently been announced by Combes and Hislop, \cite{Hislop-04}.
They have to assume that the single site distribution has a bounded density with respect to Lebesgue measure.
So far, the averaging techniques at our disposal do not seem to be sufficient enough to prove that
$\bar\xi(\lambda)$ is locally bounded for \emph{rough} single site distributions, even if they are
H\"older continuous.

In this context we would like to mention bounds on averaged fractional powers of the SSF derived in
\cite{AizenmanENSSa}.

\smallskip

%\begin{rem}
%\label{r-appl}
As a final remark, we discuss how Theorem \ref{t-SSF} can be used to improve Wegner estimates for alloy type
models with somewhat different properties than in Theorem  \ref{t-WE}.  First we present an improvement of a
recent Wegner estimate by Combes, Hislop and Klopp \cite{CombesHK-03} for single site potentials with
small support.
%\end{rem}

\begin{thm}
\label{t-CHKWE}
Let $H_\omega$ be an alloy type model as defined in the paragraph follwing \eqref{e-CHN-SSF}. Assume that $V_{\per}$ has the unique continuation property and is
bounded below, $\omega$ is distributed according to a bounded density and $0 \le u\in L^\infty$ is strictly
positive on an open set. Then for each $E_0\in \RR$ there exists a constant $C_W$ such that
\begin{equation}\label{e-CHKWE}
\EE\{\Tr \chi_{[E-\varepsilon, E+\varepsilon]} (H_\omega^l) \}
\le C_W \ \varepsilon \, \l(\log\frac{1}{\varepsilon}\right)^d \ |\Lambda|
\end{equation}
for all $E\le E_0$ and $\varepsilon \le 1/2$.
\end{thm}

This follows directly if one uses Theorem \ref{t-SSF} instead
of the $L^p$-estimates on the SSF in the Appendix of \cite{CombesHK-03}.

\smallskip

We mention three more disorder regimes where Theorem \ref{t-SSF}
may be used to simplify proofs of earlier Wegner estimates and
to improve the dependence of the estimate in the energy interval length.

\noindent (1) \emph{Single site potentials with small support and singular coupling constants.} 
Using the perturbation technique of Kirsch, Stollmann and Stolz in \cite{KirschSS-98a}, we can extend the result from
Theorem \ref{t-WE} to single site potentials $u\ge \kappa\chi_{[-s,s]^d}$ for some $\kappa,s>0$ in the case
of zero magnetic field for energies near spectral edges. In this case no assumption on the unique continuation property is needed.

\noindent
(2) \emph{Coupling constants whose distribution is continuous
merely at the extreme values.} In \cite{KirschV-02a}, Kirsch and
Veseli\'c prove a Wegner estimate for alloy type potentials with
non-positive single site potentials and coupling constants which
have merely in a neighborhood of their maximal value a continuous
distribution with bounded density. The estimate applies to
energies at the bottom of the spectrum. Using Theorem \ref{t-SSF}
in the present paper, one can improve the Wegner estimate in \cite{KirschV-02a}.

\noindent
(3) \emph{Single site potentials with changing sign}.  In \cite{HislopK-02}, Hislop and Klopp studied
alloy type models with continuous, compactly supported single site potentials, which may take values with
both signs, and bounded coupling constants which are distributed according to a bounded, piecewise absolutely
continuous density. They prove a Wegner estimate which is H\"older continuous in the energy variable and
applies to energies below the spectrum of the non-random, unperturbed operator $H_0$. The result extends to
internal spectral boundaries in the weak disorder regime. An extension of Theorem \ref{t-SSF} in the present
paper to the case where the perturbation is equal to a potential sandwiched between the square roots of the
resolvent would make it possible to improve the Wegner estimate in \cite{HislopK-02} with regards to the
energy interval length.
\smallskip

Further results on Wegner estimates for alloy type Schr\"odinger operators can be inferred from
\cite{BarbarouxCH-97b,CombesHKN-02,HupferLMW-01a,Kirsch-96,KirschSS-98b,KostrykinS-01b,
Veselic-02a,Veselic-04a} and the references therein. Let us mention specifically, that if $\mu$ has bounded
density \emph{and} $u\ge \kappa \chi_{\Lambda_1}$
the IDS is actually Lipschitz-continuous, see \cite{KotaniS-1987,CombesH-94b}.
\smallskip

Let us sketch the outline of the paper: The next section contains the definition of the SSF and the proofs of
Theorems  \ref{t-singval} and \ref{t-SSF}. In Section \ref{s-WE} we prove a lemma which is needed to deal
with singular coupling constants and complete the proof of the Wegner estimate, Theorem \ref{t-WE}.

\smallskip

\section{Bounds on the SSF\label{s-proofs}}

\subsection{Definition of the SSF\label{s-DefSSF}}

We define the SSF in three steps. Each of them extends the definition
to a larger class of operators. For proofs see, e.g., \cite{BirmanY-93}
or \cite{Yafaev-92}.

Assume first that $H_1,H_2$ are selfadjoint, lower-semibounded with
purely discrete spectrum. Then the SSF is defined as the difference
of the eigenvalue counting functions,
\[
\xi(\lambda):=
\#\{n \mid \lambda_n(H_2)\le \lambda\} - \#\{n \mid \lambda_n(H_1)\le \lambda\} ,
\]
where $\lambda_n(H)$ enumerates the spectrum of $H$, including multiplicity,
in increasing order. Consider now a pair of selfadjoint, lower-semibounded operators
such that the difference $H_2-H_1$ is trace class.
Then there is a unique function $\xi$ such that \emph{Krein's trace identity}
\begin{equation}
\label{e-KTI}
\Tr \left [ \rho (H_2)-\rho (H_1)  \right  ]
= \int \rho'(\lambda) \, \xi(\lambda, H_2,H_1) \, d\lambda
\end{equation}
holds for all $\rho \in \mathcal{C}^\infty$ with compactly supported derivative 
(actually, $\rho$ can be taken to lie in a certain Besov space,
see \cite{Peller-90}). If the operators have discrete spectrum, this
definition of  $\xi$ coincides with the previous one. It can be recovered
choosing a sequence of $\rho_\varepsilon$ which converges to a step function
as $\varepsilon\to 0$.

Finally, we weaken the trace class assumption on the operator difference.
Let $g\colon \RR \to [0,\infty)$ be a monotone, smooth function such that
$g(H_2)-g(H_1)$  is trace class. Assume that $g$ is bounded on the spectra
of $H_1$ and $H_2$. Then the SSF for the operator pair $g(H_1),g(H_2)$  is
well defined and we may set
\begin{equation}
\label{e-IP}
\xi(\lambda, H_2, H_1)
:= %\ \sign \big(g'(\lambda)\big) 
\sign(g')\ \xi\big(g(\lambda), g(H_2), g(H_1)\big) .
\end{equation}
This definition is independent of the choice of $g$. Formula \eqref{e-IP}
is called the \emph{invariance principle}.
This last definition will be sufficiently general to cover the
Schr\"odinger operators we are considering.
In the sequel we will choose $g(x)=e^{-x}$.

Alternatively, the SSF can be defined via the perturbation determinant
from scattering theory by
\[
\xi(\lambda, H_1, H_2)
:= \frac{1}{\pi}
\lim_{\varepsilon\searrow 0}
\arg \det \big[1+(H_1-H_2)(H_2-\lambda -i\varepsilon)^{-1})\big]
\]
if $(H_1-H_2)(H_2+i)^{-1}$ is trace class.

\subsection{Decay of singular values}
Weyl's asymptotic law gives the asymptotic behaviour  of the $n^{\text{th}}$
eigenvalue of the Laplacian on an open ball $B$ for large $n$. The following
simple lemma provides a robust lower bound, very much in the spirit of Weyl's law,
but valid for all $n$ and for general magnetic Schr\"odinger operators. It is the
starting point for our proof of Theorem~\ref{t-singval}.

As it costs us nothing in clarity, the lemma will be presented under
weaker hypotheses than those described in the introduction.

\begin{lem}
\label{l-Weyl} Let $H = H_A + V = (-i\nabla-A)^2 +V$ as in the introduction (cf. \eqref{E:Hdefn}),
except that we now require that $V_-$ is merely $-\Delta$ bounded with relative bound $\delta <1$.
Furthermore, let $H^\mathcal{U}$ be the Dirichlet restriction of $H$ to an arbitrary open set
$\mathcal{U}$ with finite volume $\vert\mathcal{U}\vert$
(also defined via the corresponding quadratic forms).
Then, for some constant $C$, the $n^{\text{th}}$ eigenvalue of
$H^\mathcal{U}=H_A+W$ satisfies
\begin{equation}
\label{e-Weyl}
E_n \ge
\frac{2\pi(1-\delta) d}{e} \,
\Big(\frac{n}{|\mathcal{U}|}\Big)^{2/d} \!\!-C \qquad \text{ for all }n\in\NN .
\end{equation}
\end{lem}

\begin{proof} Since the Dirichlet Sobolev space $H_0^1(\mathcal{U})$ is a
natural subset of $H^1(\RR^d)$, $V_-$ is also relatively form bounded
w.r.t.\ $-\Delta^\mathcal{U}$, the Dirichlet Laplacian on $\mathcal{U}$ with
relative bound $\delta$. The diamagnetic inequality, \cite{Simon-79c},
then implies that $V_-$ is also relative form bounded w.r.t.\ to the
Dirichlet restriction of $H_A$ to $\mathcal{U}$. That is, there exists a
$C\in\RR$ such that, as quadratic forms,
\begin{displaymath}
V_- \le \delta H_A + C.
\end{displaymath}
In particular, since $V_+$ is non-negative,
\begin{displaymath}
H^\mathcal{U}\ge H_A - V_- \ge (1-\delta) H_A - C,
\end{displaymath}
which implies the bound
\begin{equation*}
\begin{split}
\Tr (e^{-2t H^\mathcal{U}})
& \le
e^{2tC} \Tr (e^{-2t(1-\delta)H_A})
=  e^{2tC} \|e^{-t(1-\delta) H_A} \|_{\rm HS}^2\\
&=  e^{2tC}  \iint_{\mathcal{U}\times \mathcal{U}} |e^{-t(1-\delta) H_A} (x,y)|^2 dx\, dy ,
\end{split}
\end{equation*}
where $\|\cdot\|_{\rm HS}$ denotes the Hilbert-Schmidt norm.
Again, using the diamagnetic inequality for the Schr\"odinger semigroup,
e.g., \cite{Simon-79c, HundertmarkS-04}, one has the pointwise bound
$|e^{-t(1-\delta) H_A} (x,y)|\le e^{t(1-\delta) \Delta^\mathcal{U}} (x,y)$.
In particular,
\begin{displaymath}
\|e^{-t(1-\delta) H_A} \|_{\text{HS}}^2 \le
\| e^{t(1-\delta) \Delta^\mathcal{U}} \|_{\text{HS}}^2
=
\Tr ( e^{2t(1-\delta) \Delta^\mathcal{U}} )
\le
|\mathcal{U}| \, \big(8\pi t(1-\delta)\big)^{-d/2} .
\end{displaymath}
In the last line we used the fact that the kernel of the Dirichlet semigroup
$e^{2t\beta\Delta^\mathcal{U}}$ on the diagonal is bounded by the free kernel, i.e.,
\[
e^{\beta\Delta^\mathcal{U}}(x,x) \le e^{\beta\Delta}(x,x)= (4\pi t\beta)^{-d/2}
\quad \text{ for all } \beta>0 \text{ and }x\in\mathcal{U},
\]
which follows immediately from the probabilistic representation of the Dirichlet semigroup
\cite{Bass-95,Simon-78}. Thus
\begin{displaymath}
\Tr (e^{-2t H^\mathcal{U}})
\le
|\mathcal{U}| \, \big(8\pi t(1-\delta)\big)^{-d/2}.
\end{displaymath}

Let $\cN^\mathcal{U}(E)$ be the number of eigenvalues of $H^\mathcal{U}$
smaller or equal to $E$. By \v Ceby\v sev's inequality and the above bound,
\begin{displaymath}
\begin{split}
\cN^\mathcal{U}(E)
&\le
e^{2tE} \int_{-\infty}^{E} e^{2ts} d\cN^\mathcal{U}(s)
\le  e^{2tE} \Tr (e^{-2tH^\mathcal{U}})\\
&\le
|\mathcal{U}| \, (8\pi (1-\delta) )^{-d/2}  \,  t^{-d/2} e^{2t(E+C)}
= |\mathcal{U}| \, \Big(\frac{e(E+C)}{2 \pi (1-\delta) d}\Big)^{d/2}
\end{split}
\end{displaymath}
where, in the last equality, we choose $t:=\frac{d}{4(E+C)} $.
Since $n\le \cN^{\mathcal{U}}(E_n)$, this, in turn, implies the lower
bound
\[
E_n \ge \frac{2\pi (1-\delta) d}{e}  \l(\frac{n}{ \vert\mathcal{U}\vert} \right )^{2/d} -C
\]
on the eigenvalues.
\end{proof}

\begin{proof}[Proof of Theorem \ref{t-singval}]
We give the proof for $V_{\eff}$, the adaption to $V_{\eff}^l$ requires
only minor changes.  We will use the symbols $c$ and $C$ for constants
that vary from line to line; however, their dependence on $H_1$ and $H_2$
will always be as stated in the Theorem.

Without loss of generality, we can assume that the origin is contained
in the support of $u$.  We will estimate the $n^{\text{th}}$ singular
value by Dirichlet decoupling at an $n$-dependent radius $R$.  To this
end, let $R$ be sufficiently large that $\supp(u)$ is contained strictly
inside the ball of radius $R$ centered at the origin, which we will denote
by $B_R$.

Let $H_j^R$ ($j=1$ or $2$) be the Dirichlet restriction of $H_j$ to the $B_{R}$, and let
\begin{equation}
\label{e-An}
A_R := e^{-H_2^R} -e^{-H_1^R}\quad\text{and}\quad D_R:= V_{\eff} - A_R.
\end{equation}
As any Kato-class potential is relatively form bounded with respect to the Laplacian
with relative bound zero, we may apply Lemma~\ref{l-Weyl} to deduce that
$\mu_n(e^{-H_j^n}) \le C \exp(- c n^{2/d} R^{-2})$ for both $j=1$ and $j=2$.
Since $A_R$ is the difference of two \emph{non-negative} operators by the min-max theorem 
its singular values  obey the same type of bound:
\begin{equation}\label{estA}
\mu_n(A_R) \leq C \exp(- c n^{2/d} R^{-2}).
\end{equation}

If $D_n$ is bounded, then $\mu_n(V_{\eff})\le \mu_n(A_R) + \|D_n\|$.  
We now proceed to estimate the norm of $D_n$
by using the Feynman-Kac-It\^{o} formula for
magnetic Schr\"odinger semigroups with Dirichlet boundary conditions,
see \cite{BroderixHL, Simon-79c}.

Let $\Erw$ and $\Prb$ denote the expectation and probability for a Brownian motion,
$b_t$ starting at $x$.  If $\tau_R=\inf\{t>0\vert b_t\not\in B_{R_n}\}$ denotes the
exit time from the ball $B_{R}$, then
\[
(D_n f)(x)
= \Erw \l[
    e^{-iS_A (b)}\big(e^{-\int_0^1 (V+u)(b_s) ds}- e^{-\int_0^1 V(b_s) ds}\big)
    \chi_{\{\tau_n \le 1\}}(b) f(b_1) \right]
\]
where $S_A^t$ is real valued stochastic process corresponding to the purely magnetic part
of the Schr\"odinger operator. To be precise, one has to fix a suitable gauge, e.g.,
Coulomb gauge, i.e., $\mathrm{div}A=0$, for this and then use gauge invariance for the
general case, see \cite{Leinfelder}.

By taking the modulus and using the triangle inequality, one sees that the magnetic vector potential drops out:
\[
\vert D_n f\vert(x)
\le \Erw \big[
    e^{-\int_0^1 V(b_s) ds} \big\vert e^{-\int_0^1 u(b_s) ds}- 1 \big\vert
    \chi_{\{\tau_n \le 1\}}(b) \vert f(b_1)\vert \big] .
\]
Moreover, only Brownian paths which both visit $\supp u$ and leave $B_{R}$ within one unit of time
contribute to the expectation. Thus if $\tau_u$ is the hitting time for $\supp(u)$ and
$\cB=\{\tau_{R} \le 1 , \tau_{u} \le 1  \} $,
\[
\vert D_n f\vert(x)
\le \Erw \big[
    e^{-\int_0^1 V(b_s) ds} \big\vert e^{-\int_0^1 u(b_s) ds}- 1 \big\vert
    \chi_{\cB}(b) \vert f(b_1)\vert \big],
\]
so by applying H\"older's inequality,
\[
\begin{split}
\vert D_n f\vert(x)
\le & \Big(\Erw \big[
    e^{-8\int_0^1 V(b_s) ds}\big]\Big)^{1/8}
    \Big(\Erw \big[
    \big\vert e^{-\int_0^1 u(b_s) ds}- 1 \big\vert^8 \Big)^{1/8} \\
    & \Big(\Erw \big[
        \chi_{\cB}(b) \big]\Big)^{1/4}
    \Big(\Erw \big[ \vert f(b_1)\vert^2 \big]\Big)^{1/2}.
\end{split}
\]
By Kashminskii's lemma, the Kato-class condition on $V_-$ and $u_-$ implies that
the first two terms are bounded uniformly in $x$, \cite{AizenmanS,Simon-82}.

Levy's inequality combined with elementary estimates imply
$\mathbf{P}_{x=0}\{\tau_{R} \le 1 \}\le 2 \mathbf{P}_{x=0}\{|b_1|\ge R \}\leq C  e^{ - R^2/4}$.  
As any path in $\cB$ must cover the distance $r$ between $\supp  \, u$ and the complement of the ball $B_R$, we can deduce that  
$\Prb (\cB)\le C e^{-r^2/4} \le C e^{-R^2/8} $ where we chose without loss of generality $r \ge R /\sqrt{2}$. 
Thus
\[
\vert D_n f\vert(x)
\le
C e^{-R^2/32} \bigl\{ \Erw \vert f(b_1)\vert^2 \bigr\}^{1/2}
=
C e^{-R^2/32} \bigl\{ (e^{\Delta }\vert f\vert^2)(x) \bigr\}^{1/2} ,
\]
in particular, using the fact that $e^{\Delta}$ is an $L^1$ contraction,
\[
\Vert D_n f\Vert_2
\le
C e^{-R^2/32} \bigl\Vert (e^{\Delta }\vert f\vert^2) \bigr\Vert_1^{1/2}
\le
C e^{-R^2/32} \Vert f^2 \Vert_1^{1/2}
=
C e^{-R^2/32} \Vert f \Vert_2.
\]

To balance the two bounds obtained for $\mu_n(A_R)$ and $\|D_n\|$ one chooses $R:= n^{1/2d}$,
which leads to \eqref{e-singval}.
\end{proof}

\subsection{Singular value decay implies SSF estimate}
\begin{proof}[Proof of Theorem \ref{t-SSF}] Let $F_t$ and the two Schr\"odinger
operators $H_2=H_1+u$ be as in the Theorem. Using the invariance principle
and a change of variables, we have
\begin{displaymath}
\begin{split}
\int_{-\infty}^T F(|\xi(\lambda, H_2, H_1)|)\, d\lambda
&=
\int_{-\infty}^T F(|\xi(e^{-\lambda}, e^{-H_2}, e^{-H_1})|)\, d\lambda \\
&\le
e^T \int_{e^{-T}}^\infty F(|\xi(s , e^{-H_2}, e^{-H_1})|)\, d s
\end{split}
\end{displaymath}

By an estimate of Hundertmark and Simon \cite{HundertmarkS-02}, the integral on the
RHS is bounded by
\[
\begin{split}
\int_{-\infty}^\infty F(|\xi(s , e^{-H_2}, e^{-H_1})|)\, d s
&\le
\sum_{n =1}^\infty \mu_n(V_{\eff}) (F(n)-F(n-1)) \\
& \le
\sum_{n =1}^\infty \mu_n(V_{\eff}) \int_{n-1}^{n} (e^{ts^{1/d}}-1) ds
\le
C\sum_{n =1}^\infty e^{(t-c)n^{1/d}}
\end{split}
\]
which is finite, if we chose $t$ smaller than the constant $c$ from Theorem \ref{t-singval}.
This proves \eqref{e-integral}.

To prove \eqref{e-dual}, we dualize the bound \eqref{e-integral} with the
help of Young's inequality for an appropriate pair of functions.
Note that $F_t$ is non-negative, convex with $F_t'(0)=0$ and hence its
Legendre transform $G$ is well defined and satisfies
\[
G(y) := \sup_{x\ge 0} \{xy-F(x)  \}
\le y \Big ( \frac{\log(1+y)}{t} \Big)^d  \text{ for all } y \ge 0
\]
Thus, by the very definition of $G$, Young's inequality holds: $ yx \le F(x) + G(y)$.
So, with $b=\sup\supp(f)$,
\begin{equation}
\label{e-Y}
\int f(\lambda) \xi(\lambda)\, d\lambda  \le
\int_{-\infty}^b F(|\xi(\lambda)|) \, d\lambda + \int G(|f(\lambda)|) \, d\lambda
\end{equation}
Using the estimate \eqref{e-integral}, the first integral is bounded by $K_1 e^b$.
For the second integral in \eqref{e-Y}, we estimate
\[
\int G(|f(\lambda)|) \, d\lambda
\le
\int |f(\lambda)| \Big( \frac{\log(1+\vert f(\lambda)\vert)}{t} \Big)^d \, d\lambda
\le
t^{-d} |\log(1+\|f\|_\infty)|^d \, \|f\|_1 .
\]
This finishes the proof of Theorem \ref{t-SSF}.
\end{proof}

\section{Proof of the Wegner estimate\label{s-WE}}
\subsection{A partial integration formula for singular distributions}
The main new idea to deal with single site distributions that are not
absolutely continuous, is the following simple

\begin{lem}
\label{l-sme}
Let $\mu$ be a probability measure with support in $(a,b))$
{\rm (}or $(a,\infty)$, if its support is unbounded from above{\rm })
and $\phi\in C^1(\RR)$ be non-decreasing and bounded.
Then for any $\varepsilon>0$,
\[
\int_\RR [\phi(\lambda+\varepsilon)-\phi(\lambda)] \, d\mu(\lambda)
\leq s(\mu, \varepsilon)\cdot [\phi(b+\varepsilon)-\phi(a)]
\]
where $s(\mu,\varepsilon)$, the modulus of continuity of $\mu$, is
defined in \eqref{definition-s-mu-epsilon}.(If $b=\infty$, $\phi(b+\epsilon)$ means $\lim_{x \to \infty} \phi(x)$, which exists by the properties of $\phi$.)
\end{lem}

\begin{proof} The proof of this lemma follows immediately from the
well-known integration-by-parts formula for Stieltjes integrals.
We include it for the convenience of the reader. We write $d\mu=dM$, where $M$ is the
distribution function of $\mu$.  In the following, all integrals are defined as
Stieltjes integrals. Shifting variables and using that $M$ is constant outside of $[a,b]$ gives
\begin{displaymath}
\int  [\phi(\lambda+\varepsilon)-\phi(\lambda)] \, d\mu(\lambda)
=
\int_{a}^{b+\varepsilon} \phi(\lambda)  \, d[M(\lambda-\varepsilon)-M(\lambda)].
\end{displaymath}
Integrating by parts gives
\begin{displaymath}
\begin{split}
\int  [\phi(\lambda+\varepsilon)-\phi(\lambda)] \, d\mu(\lambda)
= &
\,\, \Big[\phi(\lambda)[M(\lambda-\varepsilon)-M(\lambda)]\Big]_{a}^{b+\varepsilon} \\
& - \int_{a}^{b+\varepsilon} \phi'(\lambda)[M(\lambda-\varepsilon)-M(\lambda)]  \, d\lambda .
\end{split}
\end{displaymath}
The first term is zero, since $M$ is constant outside of $[a,b]$ (in case $b=\infty$,
one uses boundedness of $\phi$ and
$\lim_{\lambda\to\infty} [M(\lambda-\varepsilon)-M(\lambda)] =0$).
The second term is bounded by
\begin{displaymath}
\begin{split}
\int_{a}^{b+\varepsilon} \phi'(\lambda)[M(\lambda)- M(\lambda-\varepsilon)]  \, d\lambda
&\le
\sup_{\lambda} \, [M(\lambda)-M(\lambda-\varepsilon)]\cdot
\int_{a}^{b+\varepsilon} \phi'(\lambda) \, d\lambda\\
&\le s(\mu,\varepsilon)\cdot (\phi(b+\varepsilon)-\phi(a)),
\end{split}
\end{displaymath}
since $\phi'\ge 0$.
\end{proof}

\subsection{Proof of the Wegner estimate}

Let $\rho$ be a switch function adapted to the interval $[E-\varepsilon,E+\varepsilon]$; see
the discussion preceding \eqref{e-SSF}.  Then
\[
\chi_{[E-\varepsilon,E+\varepsilon]} (x) \le \rho(x+2\varepsilon) -\rho(x-2\varepsilon)
\]
We may assume without loss of generality $\sum_k u(\cdot-k)\ge 1$.
By the mini-max principle for eigenvalues, we conclude
\[
\Tr  [\rho (H_\omega^l+ \varepsilon)] \le \Tr\Big[\rho (H_\omega^l+ \varepsilon\sum_k u(\cdot-k)) \Big] .
\]
Assume without loss of generality that $l\in \NN$. Then $\Lambda_l$ is decomposed in $L:=l^d$ unit cubes. We
enumerate the lattice sites in $\Lambda_l$ by $ k\colon \{1, \dots, L\} \to \tilde \Lambda = \Lambda \cap
\ZZ^d$, $n\mapsto k(n)$ and set
\[
W_0 \equiv 0, \quad  W_n =\sum_{m=1}^{n} u(\cdot-k(m)), \qquad n=1,2,\dots, L
\]
Thus
\begin{align}
\EE \{\Tr [\chi_{[E-\varepsilon,E+\varepsilon]} (H_\omega^l) ]\}
& \le       \nonumber
\EE \{\Tr [ \rho(H_\omega^l+2\varepsilon)-\rho(H_\omega^l-2\varepsilon)]\}
\\      \label{e-project}
& \le
\EE \{\Tr [ \rho(H_\omega^l+2\varepsilon)-\rho(H_\omega^l+2\varepsilon-4\varepsilon W_{L})]\}
\\      \nonumber
& \le
\EE \l \{\sum_{n=1}^{L}\Tr [ \rho(H_\omega^l+2\varepsilon-4\varepsilon  W_{n-1})-
\rho(H_\omega^l+2\varepsilon-4\varepsilon  W_{n})] \right\}
\end{align}
We fix $n \in \{1, \dots, L\}$, denote  $k_0=k(n)$,
\[
\omega^\perp := \{\omega_k^\perp\}_{k \in \tilde \Lambda}, \qquad
\omega_k^\perp :=\begin{cases} 0 \quad &\text{if $k=k_0$}, \\
\omega_k \quad &\text{if $k\neq k_0$}, \end{cases}
\]
and set
\[
\phi(\omega_{k_0}) = \Tr\bigl[\rho(H_{\omega^\perp}^l-2\varepsilon +4\varepsilon W_{n-1}+\omega_{k_0}
u(\cdot- {k_0}))\bigr], \quad \omega_n\in\RR.
\]
The function $\phi$ is continuously differentiable, monotone increasing and bounded.
By definition of $\phi$,
\[
\EE \{\Tr [ \rho(H_\omega^l+2\varepsilon-4\varepsilon  W_n))
-\rho(H_\omega^l+2\varepsilon-4\varepsilon  W_{n+1})]\}
\le
\EE \{\int [\phi(\omega_{k_0}+2\varepsilon)-\phi(\omega_{k_0})]\, d\mu(\omega_{k_0})\}
\]
Let $a=\inf\supp(\mu)-1$ and $b= \sup\supp(\mu)+1$.
Using Lemma \ref{l-sme} and the Krein trace identity \eqref{e-KTI}
together with the second part of Theorem \ref{t-SSF},  we have
\begin{align*}
\int [\phi(\omega_{k_0}+2\varepsilon)-\phi(\omega_{k_0})]\, d\mu(\omega_{k_0})
&\le
s(\mu,2\varepsilon) [\phi(b+2\varepsilon)-\phi(a)] \le C_E \, s(\mu,2\varepsilon) \l( \log(1/\varepsilon)\right)^d
\end{align*}
which implies that \eqref{e-project} is bounded by
\begin{align*}
C_E \sum_{n=1}^{L} s(\mu,2\varepsilon) \l( \log(1/\varepsilon)\right)^d
\le
C_E  \, s(\mu,2\varepsilon) \l( \log(1/\varepsilon)\right)^d \, l^d
\end{align*}
Note that we apply Theorem \ref{t-SSF} successively $L$ times. However, the constant $C_E$ depends only on
the diameter of $u$ and a local norm of the negative part of the background potential. For this local norm
exist an uniform estimate independent of $\Lambda_l$ and the configuration of the coupling constants
$\omega_k, k\neq k_0$.
\bigskip

\noindent
{\textbf{ Acknowledgements:} Stimulating discussions with W.~Kirsch, V.~Kostrykin, G.~Stolz and
S.~Warzel are gratefully acknowledged. P.S.~acknowlegdes the kind invitation to the University of Tokyo where
part of this work started. D.H and I.V.~would like to thank the Department of Mathematics at CalTech, 
especially B.~Simon and C.~Galvez, for their warm hospitality. 
We thank the following institutions for financial support: the National Science Foundation for their support under grants DMS--0400940 (D.H.) and DMS--0401277 (R.K.), the Sloan Foundation (R.K.), the JSPS through Grant-in-Aid for Scientific Research (C) 13640155 (S.N.), the DFG through the SFB 393 (P.S.) and grants Ve 253/1 and /2 (I.V.). }

\def\cprime{$'$}

\end{document}